# Selection Policy: Fighting against Filter Effect in Network of Caches

Saeid Montazeri Shahtouri
saeid@comp.nus.edu.sg
National University of Singapore

Richard T. B. Ma
tbma@comp.nus.edu.sg
National University of Singapore

*Abstract*—Many Information Centric Networking (ICN) proposals use a network of caches to bring the contents closer to the consumers, reduce the load on producers and decrease the unnecessary retransmission for ISPs. Nevertheless, the existing cache management scheme for the network of caches obtain poor performance. The main reason for performance degradation in a network of caches is the filter effect of the replacement policy. A cache serves the requests that generate cache-hits and forwards the requests that generate cache-misses. This filtering changes the pattern of requests and leads to decreased hit ratios in the subsequent caches. In this paper, we propose a coordinated caching scheme to solve the filter effect problem by introducing the selection policy. This policy manages a cache such that: i) the cache obtains a high hit ratio ii) the missed requests from the cache can be used by subsequent caches to obtain a high hit ratio. Our coordinated selection scheme achieves an overall hit ratio of a network of caches equivalent to that of edge routers with big caches. Moreover, our scheme decreases the average number of evictions per cache slot by four order of magnitude compared to the LRU universal caching.

## I. INTRODUCTION

Recently Information Centric Networking (ICN) has gained attention as one the future Internet architectures. Although there are different proposals for ICN such as NDN [14], NetInf [2] and DONA [16], all of the existing ICN proposals introduce *in-network caching*. Through in-network caching, each router uses its own cache to store data packets passing through. This leads to a *network of caches* with poor performance in terms of overall hit ratio. The main reason for this poor performance is the *filter effect* [26] or *trickle-down effect* [11].

A cache can be considered a filter. That is, the cache serves the requests that generate cache-hits and forwards the requests that generate cache-misses. This filtering changes the pattern of requests such that subsequent caches are unable to obtain high hit ratios from the forwarded requests. To reduce the filter effect, Busari and Williamson [26] proposed heterogeneous replacement policies. Later, Ari *et al.* [3] proposed Adaptive Caching using Multiple Experts (ACME) which uses neural network to find the optimal combination of replacement policies. Even though the previous studies combined different replacement policies to obtain higher hit ratio in the core routers, their results show that the filter effect still appears. This is because the request pattern in the core routers is changed by the filtering in the edge routers. Consequently, the overall hit ratio of the network of caches is degraded. The question we are addressing is how to manage caches to obtain high hit ratio in both edge and core routers.

The first contribution of this paper is a new cache management policy called *selection policy*, in which the cache fetches the content for the first $c$ (number of cache slots) different requests. Then, the cache slots are frozen for a period of time until they become stale and they can be replaced. This policy achieves two goals: i) obtains a high hit ratio ii) reduces the filter effect. Using *selection policy* instead of replacement policy at the edge routers leads to higher hit ratios in the core routers. Furthermore, we mathematically prove that the *selection policy* and the Least Recently Used (LRU) replacement policy have the same hit ratios under the Independent Reference Model (IRM). Moreover, using extensive simulations, we show that they have similar hit ratios under non-IRM. The hit ratio of the cache is further improved compared to LRU by modifying the *selection policy* to fetch $c$ content among more than the $c$ first different requested content.

Our second contribution is a proposal and evaluation of a *coordinated selection scheme*, based on the modified version of the *selection policy*. This scheme has high hit ratio, low communication overhead and works in line-speed [4]. Using simulation for both synthetic and real network topologies, we show that the overall hit ratio of a network of caches with $c$ slots is equivalent to that of a network in which only edge routers have LRU caches with size $c \times x$, where $x$ is the average number of routers (caches) between consumers and producers in the network (equal to average hop distance minus 1). In addition, our scheme obtains high hit ratio in both edge and core routers with an improvement of two times higher overall hit ratio for small cache sizes (0.05-0.5% of all data) which are important in ICN and up to 14% for large cache sizes (0.5-10% of all data) compared to LRU universal caching. Moreover, our scheme saves energy by decreasing the average number of evictions per cache slot by four order of magnitude compared to LRU universal caching scheme (LRU in all caches). Our simple implementation of our scheme does not need additional overhead information such as content popularity [20] and neighbour cache information [13]. The coordination among ICN routers is simply done by piggybacking information through an integer field in the request and the data packets.

The paper is organized as follows. Section 2 describes the *selection policy* for one single cache. In Section 3, we describe how our *selection policy* differs with the replacement policy in terms of filter effect. We propose a coordinated selection scheme in Section 4. The evaluation of our coordinated scheme is presented in Section 5. Section 6 will represent the related works. Finally, Section 7 conclude the paper.

## II. NEW CACHE MANAGEMENT POLICY

In this section, we give an overview of our new cache policy, the *selection policy*. This policy selects content for the first $c$ distinct requests and passes the following requests for a predetermined period of time. Using a probabilistic approach, we prove that the *selection policy* has the same hit ratio as an LRU replacement policy under Independent Reference Model (IRM) assumption.

### A. Overview of Selection Policy

A cache managed by *selection policy* alternates between selecting and fetching content for requests and freezing the cache slots. Figure 1 shows the state transition diagram for our *selection policy*. In *selecting* state a cache of size $c$ selects and fetches content for the first $c$ incoming distinct requests. Then, the content of the cache is *frozen* for a time interval. During this *frozen period* the cache does not replace the selected content and the incoming missed requests and content are forwarded. After this period, the cache returns to selecting state and a new selecting process takes place.

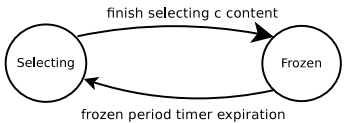

Fig. 1: Selection policy state transition diagram

The *selection policy* performance depends on the *selecting algorithm* used to populate the cache and the frozen timer. A selecting algorithm defines the way the cache entries are populated with content and this algorithm is essential for obtaining high hit ratio. Currently, the best hit ratio is obtained using a selecting algorithm based on content popularity [24]. This algorithm, used in *static caching*, requires real-time measuring of the access frequency for all content. However, this overhead might make static caching impractical in ICN routers because they have to operate in line-speed. Instead, our simple selecting algorithm to fetch different content for the first $c$ incoming requests can be implemented in line-speed.

The second aspect for achieving good performance in the *selection policy*, the frozen period, plays an important role in adapting the cache to the network dynamics, such as popularity changes and Internet path changes. A cache adapts to network dynamics only during selecting state. Hence, the *selection policy* adapts to changes only when the frozen period is smaller than the time between network changes. For example, the popularity of VoD is almost unchanged for a day [9]. and the majority (around 2/3's) of Internet paths does not change for days [18]. Hence, one might consider that the time between changes is in the order of hours.

### B. Hit Ratio of Selection Policy

Cache policies are usually evaluated based on their hit ratio [5]. Next, we prove that the *selection policy* has the same hit ratio as LRU replacement policy under IRM assumption.

**Theorem 1.** *Under IRM assumption selecting the first $c$ distinct requested content in a cache, has the same hit ratio as LRU hit ratio.*

*Proof:* Consider a set $N = \{1, 2, \ldots, n\}$ of $n$ different content, out of which $c$ content can be stored in a cache. Under IRM assumption, the $i^{th}$ most popular content is independently requested with probability $q_i$ which is the popularity of the $i^{th}$ content and $q_1 \geq q_2 \geq \ldots \geq q_n$. In addition, let $\vec{\sigma} = \{\sigma_1, \sigma_2, \ldots, \sigma_c\}$ be a possible state of a cache where $\sigma_i \in N$ is located in the $i^{th}$ slot. Moreover, we define $h_X$ as the hit ratio of a cache under the policy $X$ and $\pi_X(\vec{\sigma})$ as the steady state probability of $\vec{\sigma}$ under the policy $X$.
Based on [23], $\pi_{LRU}(\vec{\sigma})$ can be calculated by

$$\pi_{LRU}(\vec{\sigma}) = \prod_{i=1}^{c} \frac{q_{\sigma_i}}{1 - \sum_{j=1}^{i-1} q_{\sigma_j}} \quad (1)$$

Suppose a cache is managed by the *selection policy*. After entering the selecting state, the cache is filled with the first $c$ distinct requested content. That is, the first requested content is placed at the first slot, the second distinct requested content is placed at the second slot and so on. After placing the $c^{th}$ distinct requested content at the $c^{th}$ slot, the cache goes to the frozen state. Let $P_j^i$ denote the probability that the $i^{th}$ distinct requested content is content $j \in N$ given that the first $i-1$ distinct requested contents are $\sigma_1, \sigma_2, \ldots, \sigma_{i-1}$. For $i = 1$, $P_j^1$ is simply the probability that the first requested content after starting selection is content j. According to IRM, we have

$$P_{\sigma_1}^1 = q_{\sigma_1} \quad (2)$$

,and

$$p_{\sigma_i}^i = \frac{q_{\sigma_i}}{1 - \sum_{j=1}^{i-1} q_{\sigma_j}} \quad (3)$$

The explanation is that given the first $i-1$ distinct requested contents are $\sigma_1, \sigma_2, \ldots, \sigma_{i-1}$, the remaining contents compete to occupy the next position in the cache. We exclude the popularity of the already requested content and normalize the popularity of the remaining content to one. The probability of finding a cache in a frozen state $\vec{\sigma}$ can be calculated by the following expression.

$$\pi_{SEL}(\vec{\sigma}) = \prod_{i=1}^{c} p_{\sigma_i}^i = \prod_{i=1}^{c} \frac{q_{\sigma_i}}{1 - \sum_{j=1}^{i-1} q_{\sigma_j}} \quad (4)$$

It can be seen that $\pi_{SEL}(\vec{\sigma})$ is exactly equal to $\pi_{LRU}(\vec{\sigma})$. Hence, $h_{SEL} = h_{LRU}$ since for any policy $X$

$$h_X = \sum_{i=1}^{n} q_i \sum_{i \in \vec{\sigma}} \pi_X(\vec{\sigma}) \quad (5)$$

∎

## III. FIGHTING FILTER EFFECT

The performance of one cache is influenced mainly by the management policy. However, for a network of caches, such as the routers in the Internet, the cache performance is influenced not only by its management policy, but also by the interactions with other caches. For example, one of the interactions in a network of caches is the filter effect [26]. This filter effect lowers the overall hit ratio of a network of caches using replacement policies by serving the requests that generate cache-hits and forwarding the requests that generate cache-misses. Hence, there is little chance for core routers to achieve high hit ratios because their incoming requests are filtered by the edge routers.

This section shows that our *selection policy* reduces filter effect compared to replacement policies. Firstly, we explain the difference between filter effect induced by the replacement policy and the *selection policy*. Secondly, we use the average, minimum and maximum stack distances [19] as metrics to quantify the filter effect. Lastly, using these metrics, we compare our *selection policy* with three commonly used replacement policies. Our simulation results show that the *selection policy* is better than all evaluated replacement policies in terms of filtering effect.

### A. Types of Filter Effect

To achieve a high hit ratio, the replacement policy uses a specific property (*locality of reference*) of the requests stream. The locality of reference means that "a content just requested has a high probability of being referenced in the near future" [15]. The locality of reference determines the *potential* of achieving a high hit ratio. The stronger the locality of reference is, the more the potential of achieving high hit ratio exists. However, the locality of reference is weakened by the filter effect of the replacement policy. In contrast, the *selection policy* allows the missed requests to be efficiently served by other caches.

Although both selection and replacement policies serve some requests (hits) and forward some requests (misses), they have different types of filter effect on the pattern of requests. Despite the replacement policy that serves a *fraction* of requests (with strong locality of reference) of **all** content, the selection policy serves *all* the requests for $c$ number of **selected** content. Suppose that in Figure 2 both routers use replacement policy. Therefore, the locality of reference is valid for router1 but the locality of reference is weakened in router2 since the requests are affected by the filter effect of router1. That is, if router2 receives a request for a specific content, router2 cannot assume that it will receive another request for that specific content with a high probability in near future. Otherwise, router1 should miss two requests with strong locality of reference. This contradicts the functioning of replacement policy in router1. However, by using the *selection policy* in the router1, the locality of reference in the second router is still valid because router1 either serves all of the requests of one specific content or forwards all of the requests for that content. Therefore, the locality of reference can still be assumed by router2.

### B. Metrics

The stack distance is widely used in the literature to characterize the locality of reference [19]. The stack distance of the $j^{th}$ request ($j = 2, 3, \ldots$) for content $i$ is defined as the number of *distinct* content requested between the $j - 1^{th}$ and $j^{th}$ requests for content $i$ (undefined stack distance considered for the first request of content $i$). For example let 4, 5, 1, 3, 2, 7, 2, 3, 1, 6 be a stream of requests for content 1 to 7. The stack distance of the second request for content 1 is 3 because there are three distinct content (2, 3, 7) requested between the first and the second requests of content 1. The stack distance represents the strength of locality of reference. The smaller the stack distances of the content requests are, the stronger the locality of reference for the requests of that content is. Using the stack distance, we define three parameters to characterize the locality of reference: the minimum, maximum and average stack distances. The minimum (maximum) stack distance is defined as the smallest (largest) stack distance seen in a stream of requests. The average stack distance, $SD_{avg}$, is defined as

$$SD_{avg} = \frac{\sum_{i=0}^{n-1} SD(i) \times i}{\sum_{i=0}^{n-1} SD(i)} \quad (6)$$

where $n$ is the total number of content in a stream and $SD(i)$ is the number of occurrences of stack distance $i$ in the stream. The $SD_{avg}$ is not sufficient to characterize the locality of reference by itself. Two streams with similar $SD_{avg}$ may lead to totally different hit ratios in a cache because of their different minimum stack distances.

### C. Comparison of Selection and Replacement Policy

The replacement policy uses the locality of reference of **all** content to achieve a high hit ratio. Specifically, a replacement policy serves the requests with small stack distances targeting any content. This results in increasing the stack distance of the missed requests forwarded to the subsequent caches and the subsequent caches hardly obtain high hit ratios. Limiting the serving operation to the requests of the selected content leads to less increment in the average stack distance of the missed requests. In contrast to the replacement policy, the *selection policy* decreases the maximum stack distance and does not increase the minimum stack distance.

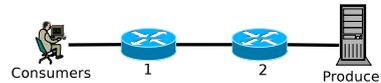

Fig. 2: Simple topology for filter effect experiment

We set an experiment with topology depicted in Figure 2 to compare the filter effect of the replacement and selection policies. In the experiment, the routers have equal cache size and there are 1000 equal sized content located on the producer. The consumers request content based on Zipf(1, 1000) distribution. We do the simulation with 15 million requests and 10 different seeds. Three replacement policies (FIFO, RND and LRU) applicable in an ICN router [4] are considered. These policies and the *selection policy* are applied as the cache management of the first router.

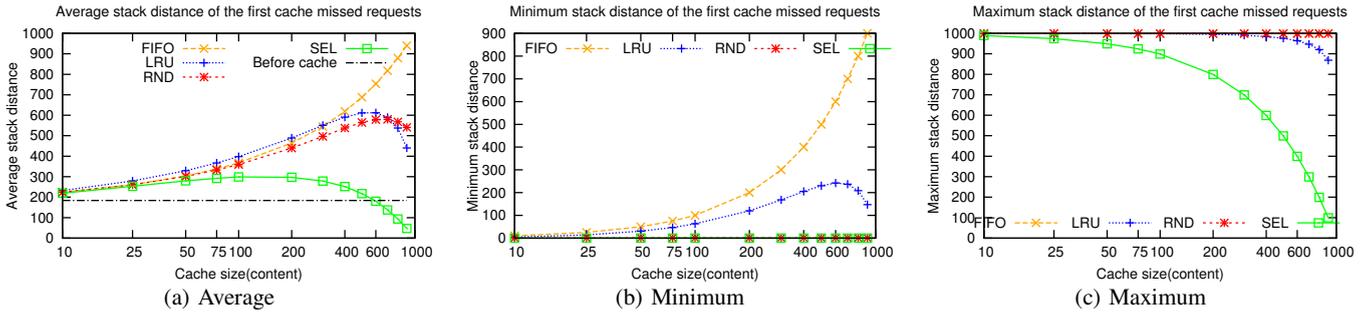

Fig. 3: Stack distances of the missed requests of the first cache

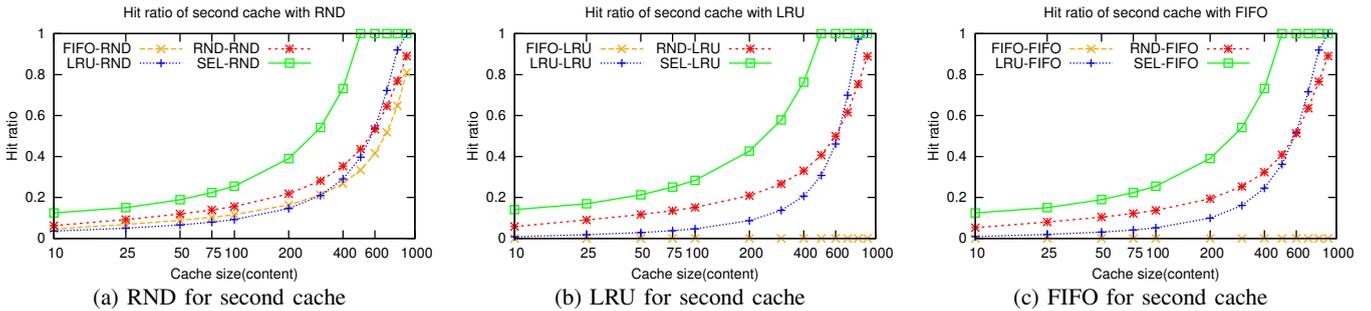

Fig. 4: Second cache hit ratio. Second cache is managed by RND(a), LRU(b) and FIFO(c). First cache is managed by RND, LRU, FIFO and selection policy

We measure the average, minimum and maximum stack distances of the first cache requests. When the *selection policy* manages the first cache, the average stack distance of the missed stream has the least increment as depicted in the Figure 3a and the minimum and the maximum stack distances have their smallest values as depicted in the Figures 3b, 3c. That is, the locality of reference of the missed requests from the *selection policy* has the highest potential to be efficiently served by another cache. Consequently, the second cache (managed by three replacement policies) has the highest hit ratio when the first cache is managed by the *selection policy* as depicted in Figure 4. A comparison between the FIFO-LRU (FIFO in router1-LRU in router2) and RND-LRU in Figure 4b shows the importance of the minimum stack distance. The hit ratio of FIFO-LRU is zero because the missed requests from FIFO has a minimum stack distance equal to its cache size. However, the hit ratio of RND-LRU is not zero because the missed requests from RND has a minimum stack distance equal to 1. It should be mentioned that FIFO and RND have very close average stack distances.

## IV. COORDINATED SELECTION SCHEME

In this section we describe our *selection policy* in a coordinated manner in network of caches. By using the *selection policy* which reduces the filter effect, both edge and core routers have the chance to obtain high hit ratios. To explain our coordinated selection scheme, we first give an overview by introducing *route-based selection priority* and *selection collision*. Next, the details of our scheme are explained. Lastly, we describe our modified *selection policy* which outperforms LRU.

### A. Overview

In our coordinated selection scheme, the caches use *selection policy* and obtain high hit ratios. However, without coordination among the caches, the overall hit ratio might degrade because of the *selection collisions*. A selection collision happens if two caches select a duplicate packet when both caches are on the same *route*. A route is defined by a four tuple of i) a group of consumers ii) a producer iii) a set of routers iv) a set of links. The sets of routers and links connect the group of consumers to the producer. For example in Figure 5, a selection collision happens in route2 (solid line) if both router2 and router3 select a common packet from producer2. The route2 connects consumers2 to producer2 through routers 2, 3, and 4.

To solve the selection collision, we propose an implicit and low overhead coordinated selection scheme. This scheme is based on two design principles that indicate that for every selection collision happens in a *route*:

1) the cache closer to the consumers has the priority to cache the duplicate packet and the farther cache should select another packet.
2) the cache farther from the consumers is responsible to detect the collision.

To decrease the average content retrieval delay for the consumers, our first design principle places the popular packets closer to the consumers. According to our principles, the farthest cache can select from the packets which have not been previously selected by the closer caches. Our second design principle makes the implementation simpler because it relieves the closer caches from detecting duplicate packets.

Being close to (far from) consumers is separately determined

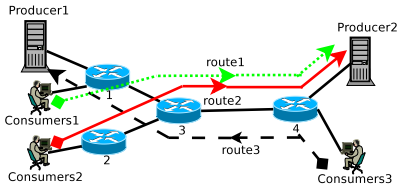
Fig. 5: General topology

based on each route. For example, router3 in Figure 5 is closer to the consumers of the route2 (consumers1) comparing to the router4. In contrast, the router3 is farther from the consumers of the route3 (consumers3) compared to the router4. Consequently, the selection priority should be *based on route* too.

*B. Route-Based Selection Priority*

Each cache (router) in the routers set of a route should have higher (lower) selection priority compared to the caches farther from (closer to) the consumers of that route. This can be achieved by using the receiver-driven communication paradigm of ICN proposals. That is, instead of selecting a data packet, a router nominates a passing request by changing the default value (-1) of a new field, Nomination Field (NF), to zero. Subsequent routers understand that the request has been nominated ($NF > -1$) and coordinate by increasing the NF value by one while the request is forwarded.

NF is increased until the request gets hit in a router or reaches a producer. In both cases, the value of the request NF is copied to the NF of the corresponding data packet. Through the reverse path towards the consumers, all intermediate routers decrease the NF value by one until the value reaches zero. The router that receives the data packet with NF equal to zero is the router which nominated the corresponding request. Only this router writes the data packet to its cache and sets the NF to $-1$. After this router in the reverse path, the NF value is not changed any more. Only one router can nominate a request when NF equal to $-1$. Using this nomination mechanism, the closest cache to the consumers has the highest selecting priority because it is the first one to receive the request.

*C. Selection Collision*

*1) Detection:* A selection collision happens when a closer router nominates a request for a specific data packet which has been previously selected by a farther router. The selection collision can be easily detected because the farther router receives a request with $NF > -1$.

*2) Resolving:* A selection collision results in a duplicate data packet which should be replaced in the farther cache. Therefore, the coordinated selection scheme needs to differentiate between duplicate data packets and other data packets. The scheme uses one extra bit called *Protection Bit* (Pb) for each cache slot to distinguish between the replaceable data packets (in *unprotected* slots with Pb=0) and non-replaceable data packets (in *protected* slots with Pb=1). The Pb of a cache slot is set after a selected data packet corresponding to a nominated request is written to that slot. If a data packet in a protected slot gets a hit for a request with $NF > -1$, the Pb of that slot is cleared.

Related to Pb, we use a variable called Unprotected Slots (US) to represent the number of unprotected slots in a cache. When a cache detects a selection collision and clears the corresponding Pb, the cache should also increase the US by one. The value of US for a sample cache is shown in Figure 6(a).

*D. Modified Selecting Algorithm*

To increase the hit ratio of the *selection policy*, the algorithm to determine how to select the data packets in the selecting state has to be modified. In Section II, we showed that if a cache selects the first $c$ requested data packets, the cache hit ratio will be similar to the LRU hit ratio. To implement the selecting algorithm which is described in Section II in our coordinated scheme, the $c$ requested data packets can be selected by two different options. Firstly, the data packets can be selected as new by *nominating* missed requests and *writing* their corresponding data packets. Secondly, the data packets can be *re-selected* as popular if the data packets from previous selection get hit after entering the selecting state. Through the first option, *nominating-writing*, a cache adapts itself to the changes of traffic pattern. By the second option, *re-selecting*, a cache selects more popular data packets. To balance between these two options, the modified selecting algorithm uses Nomination Window (NW) and Nomination Window Threshold ($NW_{th}$).

| Symb. | Meaning |
|---|---|
| NF | nomination filed: not nominated ($-1$), nominated ($\geq 0$) |
| Pb | protection bit: protected (1), unprotected (0) |
| US | # of unprotected slots |
| RS | # of remaining selections |
| NW | nomination window: # of nominated requests without receiving corresponding data packets |
| $NW_{th}$ | NW threshold |

TABLE I: Symbols in coordinated selection scheme

As summarized in Table I, NW represents the number of ongoing selections, the number of requests which have been nominated and forwarded by a router but their corresponding data packets have not been received yet. NW is used to prevent a cache from nominating requests more than it requires. The number of required selections is determined in a cache by Remaining Selection (RS). By a state transition from the frozen state to the selecting state, RS is set to the cache size ($c$). Therefore, a router can nominate $c$ requests at maximum. However, the modified selecting algorithm limits the number of concurrent nominations by $NW_{th}$. That is, the condition of $NW \leq Min(RS, NW_{th})$ should be always satisfied in a router.

If the $NW_{th}$ of a cache is set to its maximum ($c$), the cache can have $c$ ongoing selections by nominating $c$ requests before receiving any of their corresponding data packets. This leads to selecting most of the data packets through nominating-writing. In contrast, if the $NW_{th}$ is set to its minimum (1), the cache can nominate only one request and has to wait for receiving the corresponding data packet before nominating another request. This leads to providing longer time to the data packets in the unprotected slots for getting hit and being re-selected. Consequently, most of selected data packets should be from re-selecting.

*1) Nominating and Writing:* By expiring the frozen timer and entering to the selecting state, all of the data packets are considered as *stale* by clearing the Pb of all cache slots. Therefore, US and RS should be set to $c$. **Nominating:** To nominate a request two conditions should be satisfied. First, the request should not be nominated by a closer router ($NF = -1$). Second, the number of ongoing selections should be less than the number of remaining selections and $NW_{th}$ ($NW < Min(RS, NW_{th})$). If the conditions are satisfied, the cache nominate the request by setting the request NF to zero and increasing the NW by one. **Writing:** The incoming selected data packet (with $NF = 0$) should be written to an unprotected slot (with Pb=0). The unprotected slot is chosen from the first slot towards the last slot by using a variable called *Pointer* which always points to the slot that the next selected data packet should be written to. For example, a cache with two selected data packets and 6 unprotected slots is shown in Figure 6(a). The same cache is shown in Figure 6(b) after writing a selected data packet into the third slot. When the cache writes a selected data packet to an unprotected slot, the cache sets the Pb of that slot to 1, decreases the NW by one and decreases the RS and US by one.

*2) Re-selecting:* The second option for doing a selection is to re-select from the current data packets with Pb=0. These data packets are considered as either *stale* caused by entering the selecting state (by resetting all the Pbs) or *redundant* caused by a selection collision. A router re-selects a stale or redundant data packet if the data packet gets at least one hit before it is replaced. The intuition for this re-selecting is that the data packet will get referenced with high probability in near future (locality of reference). Therefore, it is a *worthwhile* data packet. For example, suppose that router3 in Figure 5 gets a hit for a data packet and the request has been nominated by router2. Then router3 clears the Pb of the corresponding cache slot and is going to replace that data packet. Router3 may get a hit for the data packet before replacing it because the data packet has not selected by router1. Therefore, continuing the caching of this data packet is worthwhile. Therefore, router3 re-select the data packet by setting its Pb, decreasing US by one and decreasing RS by one if the location of the slot is after the *Pointer* (explain why in IV-D3).

*3) Transition from Selecting to Frozen State:* A collision for a recently selected data packet may delay the cache state transition from selecting to frozen. To ensure that the transition happens after a reasonable amount of time, we differentiate between the collisions happening before and after the *Pointer*. Let us discuss the reason for this differentiation by an example. Consider the configuration of a cache in Figure 6(c). After entering the selecting state, the cache has selected four data packets located in the cache slots $1, 2, 3$ and $6$. The one in the sixth slot is selected through re-selecting. If the cache detects a collision for the data packet in the slot 2, the cache configuration is changed to Figure 6(d). Through the configuration change, the US is increased by one but the RS is not changed because selecting a new data packet for the collision which happens before the Pointer is postponed to the time that the cache is in the frozen state. Through this postponing, a cache has to quit the selecting state at most after doing $c$ selections through nominating-writing. In contrast, if the cache shown in Figure 6(c) gets a hit for a nominated request for the slot 6 both US and RS are increased by one (Figure 6(e)). It means that the selection of a new data packet instead of the redundant data packet will be done while the cache is in the selecting state. Without postponing the selections for collisions happening before pointer, a cache may stay for a long time in the selecting state. This may result in occupying many cache slots with stale data packets while the cache keeps doing new selections for the collisions. A cache state changes from selecting state to frozen state when the RS reaches zero. In the frozen state, the cache should keep doing selection until US reaches zero. In addition, if a collision happens for a cache in the frozen state, the cache should do a selection for that collision.

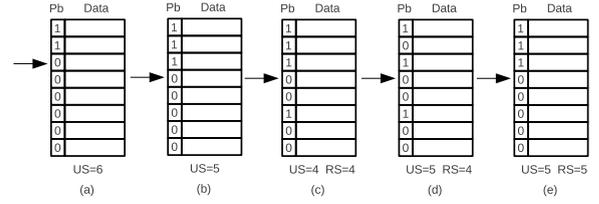

Fig. 6: Sample cache configurations

Missing a nominated request or a date packet corresponding to a nominated request may also delay or even block the state transition from selecting to frozen. It is due to the fact that the value of NW which is increased for a nomination is not decreased for writing. To solve this problem, the scheme uses a timer that timeouts if NW is not decreased for a specific period. Every time that the NW is decreased the timer is restarted. After timer timeout, the NW is divided by two. The timeout period can be set based on the average network RTT.

## V. EVALUATION

In this section, we evaluate our coordinated selection scheme from the perspective of consumers, producers and ISPs for both synthetic and real network topologies. In addition, four different types of Internet traffic with different popularity settings are used. In addition to LRU universal caching, the base for evaluation of many coordinated caching schemes [21, 8], we compare our work with a scenario called LRU-BIG which does suffer filter effect of replacement policy by having big caches only at the edge routers.

## A. Simulation setting

*1) Performance metrics:* **Producers' perspective:** i) overall hit ratio of the network of caches, $Hit_{Net}$, is defined as $Hit_{Net} = \frac{R_{entered} - R_{producer}}{R_{entered}}$, where $R_{entered}$ is the total number of requests entered to the network of cache and $R_{producer}$ is the total number of requests served by producers. **Consumers' perspective:** ii) hop reduction ratio, $H_{red}$, is defined as $H_{red} = \frac{H_{cache}}{H_{no-cache}}$ where $H_{cache}$ ($H_{no-cache}$) is the average hop distance to content with (without) caching. **ISPs' perspective:** iii) traffic reduction ratio, $T_{red}$, is defined as $T_{red} = \frac{T_{cache}}{T_{no-cache}}$ where $T_{cache}$ ($T_{no-cache}$) is the total transmitted traffic with (without) caching. It should be mentioned that the traffic transmitted between consumers and edge routers are excluded from $T_{cache}$ and $T_{no-cache}$ since caching does not affect this part of traffic. **ISPs' perspective:** iv) average number of evictions per cache slot, $E_{avg}$, is defined as $E_{avg} = \frac{E_{total}}{S_{total}}$ where $E_{total}$ is the total number of evictions in the network and $S_{total}$ is the total number of cache slots in the network. $E_{avg}$ implicitly represents the network energy consumption which is caused by the evictions in all caches.

*2) Topologies:* We consider both synthetic and real network topologies: binary tree and a US backbone topology, Abilene [1]. Our binary tree topologies, representing the aggregation, have 2, 3, 4, 5 levels of routers (3, 7, 15, 31 routers). Due to the space limit, we only present the simulation results of the largest topology which has 16 leaf routers as edge routers. Each edge router is connected to group of 125 consumers and 1 producer is connected to the tree root. In the Abilene topology, representing the core network, one edge router and one producer are connected to each core router. Each edge router is connected to a group of 100 consumers.

*3) Content popularity and size:* We use Zipf(1,1000) distribution as the content popularity distribution in the binary tree topology with only one producer. The content request rate of each group of consumers is 12.5 content/sec based on Poisson distribution because of the observation which shows the session level of Internet traffic is well modeled by a Poisson process [7]. In addition, we use four types of traffic for the Abilene topology: Zipf(0.8, 100) as web, Zipf(0.9, 100) as file sharing, Zipf(1, 100) as VoD and Zipf(1.1, 100) as UGC. There are 1100 content on 11 producers (each producer has 100 content) whose traffic type (alpha in Zipf) is randomly selected. The content request rate for each group of consumers is 22 content/sec based on Poisson distribution and each consumers group generates requests for all 11 groups of content. For both topologies, the content size is based on geometric distribution [12] with average content size of 100 packets and the packet request generation method is CBR with the rate of 100 packets/sec.

*4) Combinations of policy and cache size:* Three different combinations of policy and cache sizes are considered. The first one, LRU-EQU, is the combination of LRU universal caching and equal cache size for all routers. In addition, we consider SEL-EQU as the combination of the coordinated selection scheme and equal cache size for all routers. The last combination, LRU-BIG, combines zero cache size at the core routers and big caches at the edge routers. The size of big cache, $BIG_{size}$, for a topology is defined as $BIG_{size} = EQU_{size} \times Router_{avg}$ where $EQU_{size}$ is the cache size of combination LRU-EQU and $Router_{avg}$ is the average number of routers which a request passes to reach the content without caching (average hop distance minus one). LRU-BIG is designed to investigate that how coordinated selection scheme is successful in reducing the filter effect by using locality of reference in both edge and core routers.

*5) Other Setting:* We run simulation for 1000 seconds for 10 different seeds. The shortest path routing algorithm is used. The $NW_{th} = 1$ and the frozen period equal to 60 sec are used.

## B. Discussion-Binary Tree Topology

The overall network hit ratio in Figure 7a shows that our scheme has two times higher overall hit ratio for small cache size (0.05-0.5 %) and up to 14% higher hit ratio for large cache sizes compared to LRU universal caching. In addition, our scheme outperforms LRU-BIG configuration for the cache sizes up to point 2.5%. It is due to the fact that our scheme outperforms LRU even for the first cache (the results are omitted due to space limitation). However, for the large cache sizes the LRU-BIG is slightly better than ours because of the *worthwhile* packets described in Section IV-D2. Due to the worthwhile packets, we think that depend on the routing and topology LRU-BIG outperforms the coordinated *selection policy* in terms of overall hit ratio. However, it should be mentioned that in our simulations Big-LRU only outperforms the coordinated selection scheme for binary tree topology and cache sizes greater than 2.5% (12.5% for LRU-BIG) which is a large cache size for a router. Based on [4] an ICN router cache size can be up to 10 GB with the present memory technologies.

Although hop and traffic reduction ratios in Figures 7b and 7c show that our scheme has 4-12% higher hop reduction ratio and 3-14% higher traffic reduction ratio compared to LRU universal, LRU-BIG outperforms our scheme up to 13% which is expected due to the big caches at the edge routers. Finally, our scheme can reduce the average number of eviction per cache slot up to four order of magnitude comparing to both LRU universal and LRU-BIG. This lead to huge amount of reduction in energy consumption by ICN routers.

## C. Discussion-Abilene topology

The overall network hit ratio in Figure 7a shows that our scheme has two times higher overall network hit ratio for small cache size (0.05-0.5 %) and up to 11% higher hit ratio for large cache sizes compared to LRU universal caching. In addition, our scheme outperforms LRU-BIG configuration. The rest has the a similar trend to the binary tree topology.

## VI. RELATED WORKS

**Filter Effect:** The filter effect has been studied for several years from both frequency perspective [11, 26] and time perspective [5]. Although these works give us better understanding of the filter effect, the unique method for reducing the filter

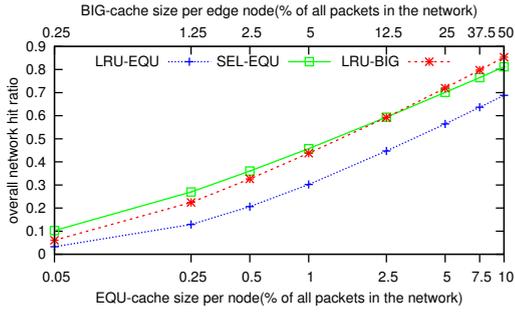

(a) Overall network hit ratio

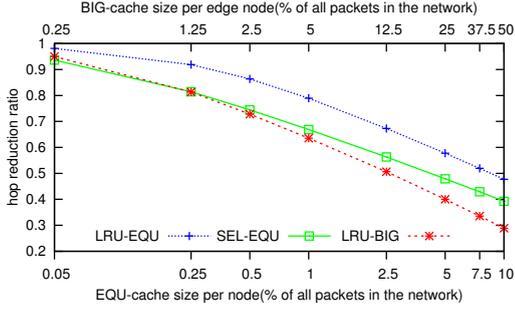

(b) Hop reduction ratio

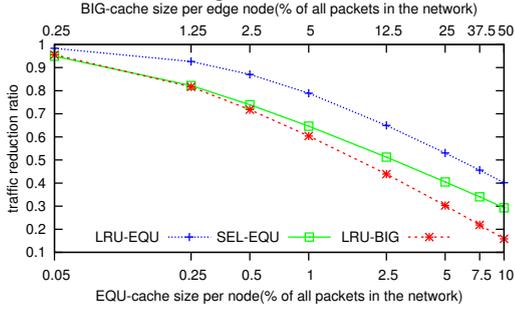

(c) Traffic reduction ratio

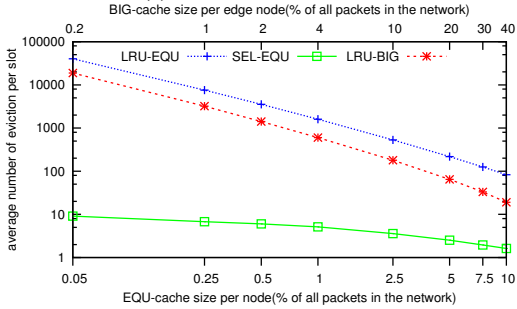

(d) Average number of eviction per cache slot

Fig. 7: Binary tree topology

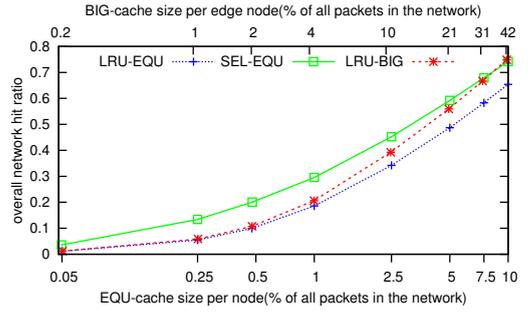

(a) Overall network hit ratio

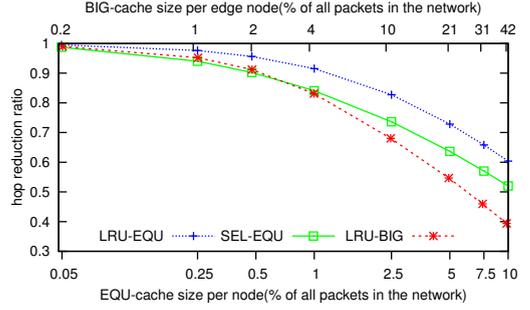

(b) Hop reduction ratio

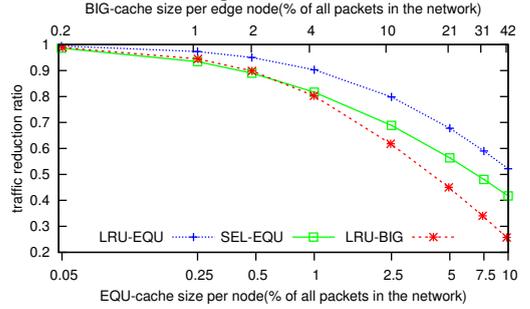

(c) Traffic reduction ratio

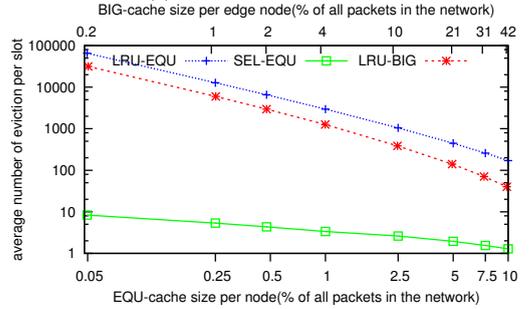

(d) Average number of eviction per cache slot

Fig. 8: Abilene topology

effect is to combine different replacement policy [26, 3], as discussed in Section I.

**Coordinated Caching Scheme:** Many works conducted on coordinated caching became impractical and/or inefficient due to the following three requirements for ICN network of caches: i) an ICN router should operate in line-speed and cannot afford high processing and communication overhead ii) the scheme has to be scalable due to the large number of routers iii) the scheme has to be applicable for general topology instead. Although previous works discuss hierarchical/distributed web cache [10] and en-route caching [6], we focus next on the coordinated caching scheme for ICN.

We divide ICN coordinated caching schemes in two categories. The first category includes [25, 13] that change the default route of requests by considering a cache's neighbours. Therefore, each cache has to maintain extra information about its neighbours' content and to periodically update its neighbours about its content changes. The updating communication overhead depends on the cache update rate which is high due to the small cache size of an ICN router compared to the total content in the Internet. The second category of the related works only coordinate through the routes from the consumers toward the producers (on-path caching). For example [22, 20, 17] impose communication and/or processing overhead: [22] needs a holistic view to decide where each content should be cached; [20] assumes that popularity of each packet is known; and [17] requires each router to measure the access frequency which imposes processing overhead to each ICN router.

Practical works such as [8, 21] proposed easy-to-implement coordinated caching schemes. WAVE, the scheme proposed by [8], determines the number of packets which should be cached by measuring the content popularity in the producers. However, measuring popularity at the producer may not be very accurate because of intermediate caches. On the other hand, the authors in [21] propose a probabilistic in-network caching scheme. The scheme considers three parameters to find the probability of writing a content in a cache: the total cache size in the path from consumer to producer, the number of hops from the previous location of the content and the number of hops to the consumer. Their idea for probabilistic caching is interesting but their evaluation is limited to the hierarchical topologies. In overall, the related works have not tackled the main cause of the poor performance in network of caches which is the filter effect.

## VII. Conclusion and Future Works

We proposed a new *selection policy* for managing a cache to achieve a hit ratio and reduce the filter effect. Using our policy, the caches receiving requests from other caches achieved high hit ratios. We showed that our *selection policy* has the same hit ratio as LRU, can be modified for higher hit ratio, and can be used in a coordinated fashion for network of caches. Our *coordinated selection scheme* argues for the use of network of caches in ICN proposals by achieving an overall hit ratio equivalent to that of a network with big caches at edge routers and no caches at core routers. Compared to LRU universal caching, our scheme obtains an improvement of two times of overall hit ratio for small cache sizes and up to $14\%$ for large cache sizes, and decreases the average number of evictions per cache slot by four order of magnitude. In addition, our scheme reduces the average hop distance to access content up to $12\%$, network traffic up to $14\%$ compared to LRU universal caching. Future works include applying new selecting algorithms with different objectives such as minimizing the Inter-ISP traffic, combining our scheme with the works that consider the content in the neighbours' caches since our scheme decreases the cache updating rate and evaluating our scheme with multi-path routing.